\documentclass [12pt,preprint]{aastex}
\usepackage{epsfig}
\begin{document}
\voffset-1cm
\newcommand{\gsim}{\hbox{\rlap{$^>$}$_\sim$}}
\newcommand{\lsim}{\hbox{\rlap{$^<$}$_\sim$}}

\title{Long gamma-ray bursts without visible supernovae: \\
a case study of redshift estimators and alleged novel objects }

\author{Shlomo Dado\altaffilmark{1}, Arnon Dar\altaffilmark{1}, A. De 
R\'ujula\altaffilmark{2} and Rainer Plaga\altaffilmark{3}}

\altaffiltext{1}{dado@phep3.technion.ac.il, arnon@physics.technion.ac.il,
dar@cern.ch.\\
Physics Department and Space Research Institute, Technion, Haifa 32000,
Israel}
\altaffiltext{2}{Alvaro.Derujula@cern.ch; Theory Unit, CERN,
1211 Geneva 23, Switzerland \\ 
Physics Department, Boston University, USA}
\altaffiltext{3}{Rainer.Plaga@gmx.de; Franzstr. 40, D-53111 Bonn, Germany}

\begin{abstract} 

It has been argued that the observational limits on a supernova (SN)
associated with GRB060614 convincingly exclude a SN akin to SN1998bw as
its originator, and provide evidence for a new class of long-duration
GRBs. We discuss this issue in the contexts of indirect `redshift
estimators' and of the fireball and cannonball models of GRBs. The latter
explains the unusual properties of GRB060614: at its debated but favoured
low redshift (0.125) they are predicted, as opposed to exceptional, if the
associated core-collapse SN is of a recently discovered, very faint type.
We take the occasion to discuss the `association' between GRBs and SNe.

\end{abstract}


\section{Introduction: the GRB/SN association}
\label{intr}

An overwhelming fraction of well observed long-duration gamma-ray bursts
(GRBs) are produced in supernova (SN) explosions.
The development of this observational conclusion on the
``SN/GRB association"  had a long and tortuous history. Within errors,
SN1998bw (Galama et al.~1998)  coincided in time and position with
GRB980425 (Pian et al.~1999).  This was widely interpreted as a chance
coincidence, or as an association between a rare type of faint GRB and a
{\it hypernova} --a member of a rare class of very energetic,
core-collapse, Type Ib/c SNe (see, e.g.~Woosley, Eastman \& Schmidt 1998,
Iwamoto et al.~1998). In the opposite extreme, Wang \& Wheeler~(1998), Cen
(1998) and Dar and Plaga (1999) argued that long GRBs are produced by core
collapse SNe akin to SN1998bw.

The majority opinion that  SN1998bw/GRB980125 was an odd couple was
maintained for a lustrum, perhaps because their association was so
difficult to accommodate within the standard views on GRBs. Yet, there
was clear photometric evidence from the optical afterglow of 
all GRBs localized at redshift $z\!<\! 1$ that the `late bumps' in their
light curves were not mere echoes. Instead, they were
all compatible with a SN1998bw template, corrected
for redshift and extinction (Dado, Dar \& De R\'ujula~2002, 2003a). Our 
understanding of the `background' to the SN (by
definition, the GRB's afterglow) was sufficiently good to 
foretell the relative contribution of an associated SN in four cases.
The most notable was GRB030329. Its first six days of afterglow data 
were sufficiently precise 
to predict the date when the SN would be bright enough to be
discovered spectroscopically (Dado, Dar \& De R\'ujula~2003b). 

The discovery, on the predicted date, of SN2003dh (Stanek et al.~2003; Hjorth et al.~2003), 
associated with GRB030329 --and with a spectrum very 
similar to that of SN1998bw-- made the majority opinion swiftly change, in 
the direction persistently advocated within the Cannonball (CB) model of 
GRBs (e.g., Dar \& Plaga~1999; Dar \& De R\'ujula~2000a; Dado, Dar \& De 
R\'ujula~2002, 2003a;  Dar \& De R\'ujula~2004; and references therein). 
Other spectroscopically proven associations with 1998bw-like SNe, such 
as GRB030213/SN2003lw (Malesani et al.~2003), strengthened the newly 
accepted credo. A paraphrase, not ``within a specific model" of Dado et 
al.~(2002) by Zeh, Klose \& Hartmann (2004) reconciled the new paradigm 
with the old data. Moreover, various GRB/SN associations indicated that 
SNe producing GRBs can differ significantly from SN1998bw (e.g., 
GRB021211/SN2002lt: Della Valle et al.~2003; GRB060218/SN2006aj: 
Campana et al.~2006; Pian et al.~2006; Mazzali et al.~2006).

Recently three different groups (Gal-Yam et al.~2006; Della 
Valle et al.~2006; Fynbo et al.~2006) reported on their failure 
to detect a SN to a deep limit in the optical afterglow (AG) of the nearby 
($z\!=\!0.125$) long-duration ($\sim\!100$ s) GRB 060614 (Parsons et al.~2006; 
Golenetskii et al.~2006). They concluded that this, as well as 
the combination of a long duration but a short temporal lag 
between different energy bands (e.g. Gehrels et al.~2006) provide 
evidence for a 
new class of long GRBs not associated with SNe. At the cited redshift,
an association with a SN as bright as SN1998bw is excluded.
And that is why this GRB was a surprise,
while a few years earlier it would have been regarded as a blessing.

The history of the theoretical ideas behind a possible GRB/SN association
is also complex and sinusoidal. Supernovae as the originators of GRBs 
were first discussed
by Colgate (1968), Goodman, Dar \& Nussinov~(1987) and Dar et al.~(1992).
The concrete realizations of the idea in these early works are obsolete,
with the possible exception of mergers of compact objects as a
mechanism behind {\it short} GRBs (Goodman et al.~1987).

 Arguing against a GRB/SN association, Woosley (1993) proposed  
a {\it failed supernova} scenario, in which
the collapse of a very massive star into a black hole would result in a GRB 
unaccompanied by a SN. Following the discovery of 
GRB980425/SN1998bw,  Woosley, Eastman \& Schmidt~(1999) and 
MacFadyen \& Woosley~(1999) proposed two
{\it collapsar} scenarios in which, instead of a `failed SN', the opposite
extreme results: a very bright and energetic hypernova
[the `delayed'  collapsar scenario involves an intermediate state
in which a neutron star is formed, as discussed in De R\'ujula
(1987) in connection with SN1987A].

An intermediate quiet period is also invoked
in the {\it supranova} model (Vietri \& Stella 1998), in which
an ordinary SN produces a neutron star. Due to loss of angular
momentum by magnetic-dipole radiation, the neutron star
collapses to a black hole and emits a GRB  years later,
when the SN is no longer observable.
Similar two-stage processes had been considered
by Shaviv \& Dar (1995) as possible mechanisms for the generation
of both short and long GRBs. In their work, the intermediate compact object
is a neutron, hyperon or quark star, and the second transition
is due to cooling, loss of angular 
momentum, mass accretion or a merger. But their suggestion
for the GRB-generating microphysics is very specific: inverse
Compton scattering, the mechanism adopted in the CB model
(e.g.~Dar \& De R\'ujula 2004).
The other scenarios for GRB/SN associations which
we have reviewed are intertwined with the
prevailing fireball models, which invoke synchrotron radiation
from relativistic thin colliding shells of delicately baryon-seeded 
shock-accelerated $e^+e^-$ pairs (e.g.~Waxman 2003).

A supranova or any other delayed-GRB origin, or a 
`failed supernova', would all explain the absence
of an observable SN associated with GRB060614, but not the presence
of a SN in all other well-observed cases.

In this paper, we examine whether or not
the deep limits on an underlying SN in the optical AG of GRB060614 
do indeed provide conclusive evidence for long GRBs which are not 
associated with SNe. We discuss and compare three alternative explanations 
why GRB060614 may have had a SN progenitor, which
was not detected:

\begin{itemize}

\item 
The GRB was produced by a very faint SN, akin to the ones discovered
by Turatto et al (1998) and by Pastorello et al.~(2004,2007),
in the outskirts of a dwarf galaxy at $z\!=\!0.125$, near the GRB's sky
position (Della-Valle et al.~2006). The brightness of the SN may have 
been further extinct below the detection limit of HST (Gal-Yam et 
al.~2006) by dust in
the host galaxy. We shall see that this possibility and the next are only
apparently incompatible with the measurements (Mangano et al.~2006).

\item
The GRB was produced in the putative host galaxy at $z\!=\!0.125$,
within a dense molecular cloud. The  EUV and soft X-rays of the GRB
destroyed the dust.
In the CB model this occurs only inside an extremely narrow cone 
along the jet axis. Through this cone only a very small fraction 
of the SN light could be seen. Most of the SN light
in the direction of the observer travelled outside this cone and 
encountered a large column density of dust, suffering strong 
extinction. 

\item 
The GRB was a normal long GRB produced in a SN explosion at a large 
redshift ($z\!\sim\! 2$) as suggested by various GRB redshift estimators 
(Schaefer \& Xiao 2006; Dado, Dar \& De R\'ujula 2007a) where from the 
SN was well below the detection 
limits. The proximity of the line of sight to GRB060614 to a foreground galaxy 
at $z\!=\!0.125$ was a chance coincidence, as advocated by Schaefer \& 
Xiao (2006) and Cobb et al.~(2006). The GRB could have been host-less,
or could have taken place in a dwarf galaxy below the detection limit
of the search by Gal-Yam et al.~(2006). 

\end{itemize}

The peak energy, equivalent isotropic energy and peak luminosity of the
GRB play a central role in our discussion of the above possibilities,
because of the relation between these observables and the luminosity of
the SN, which is explicit in the CB model (Dar \& De R\'ujula 2004). In
the first case these observables coincide with the CB model expectations
for a {\it very faint} SN at $z\!=\!0.125$. In the second case, they do not. For
the third itemized possibility, at $z\!\sim\! 2$, the values of the cited
observables are, in the CB model, the ones predicted for a conventional
1998bw-like associated SN. But they are also the ones expected from
model-independent `redshift estimators'. This gives us an occasion to
comment on their use, and on the origin of the observed correlations
between GRB observables.

\section{The CB model} 

In the CB model (Dar \& De R\'ujula~2000a, 2004; Dado et al.~2002, 2003), 
{\it long-duration} GRBs 
and their AGs are produced by bipolar jets of CBs,  ejected in 
core-collapse SN explosions (Dar \& Plaga~1999). An 
accretion disk  is hypothesized to be produced around the newly 
formed compact object, either by stellar material originally close to the 
surface of the imploding core and left behind by the explosion-generating 
outgoing shock, or by more distant stellar matter falling back after its 
passage (De R\'ujula~1987). As observed in microquasars, each 
time part of the disk falls abruptly onto the compact object, a 
pair of CBs made of {\it ordinary plasma} are emitted with high 
bulk-motion Lorentz factors, $\gamma$, in opposite directions along the 
rotation axis, where from matter has already fallen onto the compact 
object, due to lack of rotational support. 
The $\gamma$-rays of a single 
pulse in a GRB are produced as a CB coasts through the SN {\it glory} 
--the SN light scattered away from the radial direction
by the pre-SN ejecta. The electrons enclosed in the 
CB Compton up-scatter glory's photons to GRB energies. Each pulse of a GRB 
corresponds to one CB. The baryon number, Lorentz factor, 
and emission time of the individual CBs
 reflect the chaotic accretion process and are not currently predictable, 
but given these parameters (which we extract from the analysis of GRB 
AGs), all properties of the GRB pulses follow (Dar \& De R\'ujula~2004).

Two mechanisms contribute to a GRB and its afterglow: inverse
Compton scattering (ICS) and synchrotron radiation
(Dado, Dar \& De R\'ujula 2007b,
 De R\'ujula 2007). The second mechanism typically 
dominates in the AG phase. It is due to electrons of the interstellar
medium (ISM) swept-in by the CBs and spiraling 
in the their inner magnetic fields
(Dado et al.~2002, 2006) and to ISM electrons scattered to higher 
energies by the CBs and meandering in the galactic
field (Dado \& Dar 2005).
The first cited mechanism, ICS, typically
dominates the $\gamma$ and X-ray production during the `prompt'
phase. It is the one resulting in the expectations to be discussed next
 (Dar \& De R\'ujula 2004).

Let $\theta\!=\!{\cal{O}}$(1 mrad) be the typical viewing angle of an observer of a CB that
moves with a typical Lorentz factor $\gamma\!=\!{\cal{O}}(10^3)$.  Let 
$\delta\!=\!{\cal{O}}(10^3)$ be the corresponding Doppler factor:
\begin{equation}
\delta \equiv {1\over\gamma\,(1-\beta\, \cos\theta)}
                       \simeq  {2\, \gamma
                       \over 1+\gamma^2\, \theta^2}\; ,
\label{delta}
\end{equation} 
where the approximation is excellent 
for  $\theta\ll 1$ and  $\gamma \gg 1$. 
For a typical angle of incidence, the
energy of a Compton up-scattered photon from the SN glory
is Lorentz and Doppler boosted by a factor $\sim\!\gamma\,\delta/2$
and redshifted by $1\!+\! z$. The peak energy $E_p$ of the GRB's 
$\gamma$-rays
is related to the peak energy, $\epsilon_p\!\sim\! 1$ eV, of the
glory's light by:
\begin{equation}
 E_p\simeq {\gamma\,\delta\, \epsilon_p\over 2\,(1\!+\! z)}\simeq
 (250\;{\rm keV})\; {\gamma\,\delta\over 10^6}\,{2\over 1+z}\,
{\epsilon_p\over 1\;\rm eV}\; .
 \label{eobs}
\end{equation}  
The upscattered radiation,  emitted nearly isotropically 
in the CB's rest frame, is boosted by its highly relativistic motion
to a narrow  angular distribution whose number density is:
\begin{equation}
{dn_\gamma \over d\Omega}\simeq {n_\gamma \over 4\, \pi}\, \delta^2 
                       \simeq {n_\gamma \over 4\, \pi}\, {4\, \gamma^2
                       \over (1+\gamma^2\, \theta^2)^2}\, ,
\label{beaming} 
\end{equation} 
and, for a GRB of known $z$, the spherical equivalent
energy, $E_\gamma^{\rm iso}$, is  (Dar \& De R\'ujula 2004): 
\begin{equation} 
E_\gamma^{\rm iso} \simeq 
{\delta^3\, L_{_{\rm SN}}\,N_{_{\rm CB}}\,\beta_s\over 6\, c}\,
                      \sqrt{\sigma_{_{\rm T}}\, N_b\over 4\, \pi}\sim
                      (3.8\! \times\! 10^{53}\,{\rm erg})\,{\delta^3\over 10^9}\,
{L_{_{\rm SN}}\over L_{_{\rm SN}}^{\rm bw}}\,{N_{_{\rm CB}}\over 6}\,
\beta_s\sqrt{ N_b\over 10^{50}}\; ,
\label{eiso} 
\end{equation} 
where $L_{_{\rm SN}}$ is the mean SN optical luminosity just 
prior to the ejection of  CBs, $N_{_{\rm CB}}$ is the number of CBs in 
the jet, $N_b$ is their mean baryon number, $\beta_s$ is the comoving early
expansion velocity of a CB (in units of $c/\sqrt{3}$), 
and $\sigma_{_{\rm T}}$ is the Thomson cross section. The early SN luminosity required to 
produce the mean isotropic energy, $E_\gamma^{\rm iso}\!\sim\! 4\!\times\! 10^{53}$ 
erg, of ordinary long GRBs is 
$L_{_{\rm SN}}^{\rm bw}\!\simeq\! 5\!\times\! 10^{42}\, {\rm erg\, 
s^{-1}}$, the estimated early luminosity of SN1998bw. 
The observed peak isotropic luminosity, reached in the rise-time of a GRB's pulse  
($\sim\!1/2$ the time it takes a CB to become transparent to radiation)  is:
\begin{equation}
L_p^{\rm iso}\sim {\delta^4\,\beta_s^2\,L_{_{\rm SN}}\over 48\pi\,(1\!+\! z)^2}
\sim (8.3\! \times\! 10^{51}\,{\rm erg\,s^{-1})\,{\delta^4\over 10^{12}}\,
{4\, \beta_s^2\over (1+z)^2}\,{L_{_{\rm SN}}\over L_{_{\rm SN}}^{\rm bw}}}
\, .
\label{lpeak}
\end{equation}

\section{A faint SN parent of GRB060614?}
An anomalous transient in the galaxy M85, 
discovered by Kulkarni et al.~(2007) on 7 January 2006, had a very
low R-band luminosity, constant over
more than 80 days, a red colour, and narrow spectral lines. It was
identified by Pastorello et al.~(2007) as a Type II plateau SN of
extremely low luminosity, corresponding to absolute R magnitude $-12$. 
The HST data of Gal-Yam et al.~(2006b) rule out a SN brighter than 
$M_{\rm V}\!=\!-12.3$. If GRB060614 
was produced by such a faint   SN, it
would have been below the HST detection limit.   

Moreover, Mangano et al.~(2006) reported that the SWIFT WT data 
on the early X-ray AG of GRB060614 
``show strong spectral evolution with time, with average photon index 
$1.65\!\pm\! 0.04$ in the
time interval 90-270 s from the trigger and 
$2.95 \!\pm\! 0.11$ in the time
interval 270-460 s. WT spectra show evidence of
absorption at the level of $N_H\!=\!(1.3 \!\pm\! 0.3)\,10^{21}$ cm$^{-2}$,
in excess with respect to the Galactic $N_H\!=\!3\times10^{20}$ cm$^{-2}$.
The PC spectrum extracted from the second orbit of data
is well fitted by an absorbed power law with photon index
$1.8 \!\pm\! 0.2$ and $N_H$ consistent with the Galactic value."
The motion of the CBs may result in the observed decreasing absorption.
During the 90-270 s they are already
at a distance $\gamma \delta c\,t /(1\!+\!z)\!\sim\! 1$ pc from the SN, whose
radius after a day is a mere $\sim\!10^{16}$ cm (for an expansion velocity of 
$\sim\!1000$ km/s). Thus the dust column density to the SN and the 
corresponding extinction of the SN light can be much larger than that 
estimated from the WT photon spectrum during the 90-270 s interval. 
This could have dimmed the faint SN to well below the HST detectability 
limit. 

The faint SN
discovered by Turatto et al.~(1998) and the SN in M85 observed by 
Pastorello et al.~(2007) are 
$\sim\!10^{2}$ to $10^3$ times less luminous than the SNe associated with normal 
GRBs (Pastorello et al.~2004). If that is the main difference between the 
two SN types, the GRBs associated with faint SNe should have
$E_\gamma^{\rm iso}$ and $L_p^{\rm iso}$ $\sim\!10^{2}$ to $10^3$ times smaller 
than usual, see Eqs.~(\ref{eiso},\ref{lpeak}). If the initial luminosity 
of core-collapse SNe after shock break out 
is proportional to the kinetic energy of their ejecta, the faint SN in M85 
 --whose expansion velocity was $\approx$800 km/s, 
$\sim\! 20 $  times slower than that of 1998bw-like SNe-- should have had 
an 
initial luminosity $L_{_{\rm SN}}\!\sim\! 10^{40}\, {\rm erg\, s^{-1}}$,
implying $E_\gamma^{\rm iso}\!\sim\! 10^{51}$ erg, and 
$L_p^{\rm iso}\!\sim\! 10^{50}$ erg s$^{-1}$. These numbers are 
roughly consistent with the data on GRB060614: for at $z\!=\!0.125$, 
$E_\gamma^{\rm iso}\!\simeq\! 1.58^{+0.07}_{-0.13}\!\times\! 10^{51}$ erg
and $L_p^{\rm iso}\!\simeq\! 2.19^{+0.3}_{-0.6}\!\times\! 10^{50}$ erg 
s$^{-1}$,
for the standard cosmology ($\Omega\!=\!1;\, \Omega_M\!=\!0.27;\, 
\Omega_\Lambda\!=\!0.73;\, h\!=\! 0.7$).
The `downscaling' of $L_{_{\rm SN}}$ 
also explains the short lag-time, $t_{\rm lag}\!\sim\!3$ ms, of GRB060614,
emphasized by Gehrels et al.~(2006) as requiring a new class of GRBs, since
$t_{\rm lag}$ is linearly anti-correlated to peak luminosity, see
Schaefer and Xiao (2006). At $z\!=\!0.125$, Eq.~(\ref{eobs}) predicts 
$E_p\!\simeq\! 444$ keV, also
in agreement with the observed $E_p\!\simeq\! 302^{+214}_{-85}$ keV 
(Golenetskii et al.~2006). 
The above expectations for $E_p$, $E_\gamma^{\rm iso}$ and $L_p^{\rm iso}$
are shown in Figs.~(\ref{fig1},\ref{fig2}) as rectangles. The plotted (FWHM)
range of $E_p$ values is also a prediction (Dar \& De R\'ujula 2004).

Faint core-collapse SNe may produce GRBs with a very small 
$E^{\rm iso}_\gamma$ but with an ordinary $E_p$, like 060614. 
Such intrinsically faint GRBs can only be detected at relatively small
redshifts, unlike ordinary GRBs, which can be seen at larger $z$, and are
generated by bright SNe akin to SN1998bw.  Thus, although an estimated
4-5\% of core-collapse SNe are of the faint type (Pastorello et al.~2004)
they may produce only a very small fraction of the $\sim\!100$ GRBs of
known $z$. This fraction cannot be reliably estimated with the meager
information at hand.

To conclude, GRB060614 may have occurred at $z\!=\!0.125$ and be otherwise
normal, but for the low luminosity of its progenitor SN, if it was, like
the one in M85 (Pastorello et al.~2007), some three orders of magnitude
less luminous than SNe akin to SN1998bw. In the CB model this conclusion
is reinforced by the consequent predictions of the properties of the GRB,
which are correct.

\section{A GRB inside a molecular cloud?}

A molecular cloud (MC) is a region of  dense gas and dust 
($n_{_{\rm MC}}\!\simeq\! 10^3\, {\rm cm^{-3}}$) which shields its contents 
against the ambient ultraviolet radiation. In such 
a cold, protected environment, the predominant form of matter, atomic 
hydrogen, preferentially associates into molecular hydrogen. 
Star formation is presumed to begin in the cores of 
MCs,  when they 
become gravitationally unstable and  fragment into smaller 
clouds that collapse into proto-stars. The very massive 
stars evolve rapidly and end up in SNe,  which produce 
shock waves that trigger more star formation and SNe. 
The optical light from the first SNe in the
MC is strongly  extinct by the dust.  
Later, the winds from massive stars and the SN ejecta 
sweep up the ISM and eventually form a superbubble 
transparent to optical light.  
    
The radiation of GRBs is intense enough to destroy 
the dust on its way out of a  MC (Waxman \& Draine 
2001). But, in the CB model, the angular size of a GRB's beaming cone 
subtends only a small fraction $\sim 1/\gamma^2$ 
of the SN photosphere, see Eq.~(\ref{beaming}).
Most of the SN light pointing to the observer passes through 
the region of the MC lying outside the beaming cone, and
is strongly extinct by dust. Hence, while most of the beamed AG from 
 CBs is visible to an observer with a typical viewing angle 
$\theta\sim 1/\gamma$, only a fraction $\sim 1/\gamma^2$ of the 
SN light reaches the observer. This fraction is too faint to be detectable.
The decrease of the column density in front of the jet as a
function of time ---inferred from the prompt and early-time X-ray AG 
of GRB060614--- and the initially rising light-curve of its optical AG 
(Mangano et al.~2006) are consistent with the MC interpretation.

For a GRB originating in a MC, the
CB model predicts a strong extinction of the light of the 
associated SN, without a 
comparable extinction of the late GRB's AG. But it cannot explain 
without further ado, in the case of GRB060614 at $z\!=\!0.125$,
the large  $E_p$ and the small $E_\gamma^{\rm iso}$ 
and $L_p^{\rm iso}$: the first implies a typical
$\delta$, the two others favour a significantly smaller one, 
see Eqs.~(\ref{eobs},\ref{eiso},\ref{lpeak}).
This possibility suggests itself
naturally. But it is disfavoured.

\section{Correlations and red-shift estimators; a normal GRB at a typical $z$?}

GRB060604 had  `normal'  duration, fluence, spectrum, peak energy and 
energy flux, pulse widths, variability, and X-ray and optical AGs.
This suggests a redshift near the average 
for long GRBs (the mean $z$ of 40  GRBs with 
secured redshift of BeppoSAX, HETE, IPN and INTEGRAL, is 
$\bar z\!\simeq\! 1.4$; it is $\bar z\simeq 2.5$ for 45 GRBs seen by SWIFT). If 
GRB060614 originated at the average of these means 
($z_{\rm av}\!\simeq\! 1.95$) its 
proximity to a  $z\!=\!0.125$ galaxy (Price et al.~2006) 
was a  coincidence, for which Cobb et al.~(2006) estimate a 2\% 
probability (for a galaxy at 
least as bright as the putative host), consistent with $\approx$ 180 
GRBs previously detected by SWIFT.  
At  $z_{\rm av}\!\simeq\! 1.95$, the isotropic energy 
and peak luminosity of GRB060614 were also normal: 
$E_\gamma^{\rm iso}\!\simeq\! 3.7\!\times\! 
10^{53}$ erg and $L_p^{\rm iso}\!\simeq\! 1.2\!\times\! 10^{53}\, {\rm erg\, s^{-1}}$. 
Its early  X-ray AG was similar in 
magnitude, spectrum and shape to the `canonical' ones
(Nousek et al.~2006, Dado et al.~2006)  
of distant GRBs, such as  GRB 050315, also at $z\!=\! 1.95$, and with  similar 
duration, $T\!\simeq\! 90$ s. At such a redshift, 
a 1998bw-like  SN is invisible.

Gal-Yam et al.~(2006, 2006b) and Gehrels et al.~(2006) argue that $z>1$ 
is excluded for three reasons.
No Lyman-limit break in the spectrum of the AG 
of GRB 060614 was detected by the SWIFT UVOT filters. The 
probability that the line of sight to GRB 060614 passes so close to a dwarf 
foreground galaxy at $z\!=\!0.125$ is very small. There is no evidence 
from the HST spectrum obtained by Gal-Yam et al.~(2006) for any absorption 
due to dust along the line of sight in the foreground galaxy. Although these 
arguments
make $z\!\sim\! 2$ less likely, they do not exclude it: some
quasars with $1\!<\!z\!<\!2$ in the HST quasar absorption line 
key project (Jannuzi et al.~1998) show no Lyman limit breaks redward of 
1800 {\AA}. The column density of dust along the line of sight to GRB 060614 
in the foreground galaxy may be small.  A-posteriori estimates of a sky 
coincidence probability for single events are unreliable.

Schaefer and Xiao (2006) used 8 GRB redshift 
estimators (single power-law fits to correlations between pairs of GRB 
observables) to argue that GRB060604 took place at  $z\!=\!1.97\pm 
^{0.84}_{0.53}$. But since it had all the properties of normal 
GRBs, any  estimator yields for this GRB a 
redshift comparable to the mean. 
These authors argue that the estimators 
are accurate, well understood (a-posteriori) and 
predictive. But the estimators are based 
on arbitrary power laws and
the data have a large dispersion around the best 
fits. The inevitable dispersion is due to the case-by-case
variability of the parameters determining the properties 
of a GRB, whatever these {\it hidden variables} may be. 
Suppose that a 
GRB of known $z$ is an `outlier': it is relatively far from one or 
more of the mean trends of the correlations. No doubt that is due to
an atypical value of one or more hidden variables. Without a deeper
understanding, no averaging over large sets of data
and estimators would bring this GRB to the redshift
where `it should be'. An estimate
of its `best' $z$ from the estimators' mean trends would necessarily be wrong.
Often, outliers of known $z$ (typically GRB980425, but also others) are eliminated 
from the fits leading to redshift estimators. Their subsequent use to determine 
$z$ for a single debatable case is then a logical inconsistency, unless the
`misbehaviour' of the outliers is understood (like for Andromeda, at $z\!<\!0$
in Hubble's plot).

The origin of the established correlations between GRB 
properties, and the `hidden variables' responsible for their 
dispersion, are well specified in the CB model. This may help to 
asses the reliability of redshift estimators for individual GRBs.  
Most of these correlations stem from the CB-model's trivial beaming
properties, see Eqs.~(\ref{delta}-\ref{lpeak}).
They were first proposed by Shaviv \& Dar (1995) for the
$\gamma$-ray
polarization, used to predict many correlations in Dar \& De R\'ujula 
(2000b),
and shown to agree with the data in Dar \& De R\'ujula (2004).

One of the best established GRB correlations is the 
 `Amati correlation', whose latest version is $(1\!+\! z) E_p\!\simeq\! 
77\!\times\! 
(E_\gamma^{\rm iso}/10^{52}\,{\rm erg})^{0.57}$ keV. But the observed 
values of $\log[(1\!+\! z)E_p]$ are spread around the central fit
by $\pm 0.4$ (Amati et al.~2006) implying that the correlation yields a 
poorly determined redshift with $\Delta [\log (1\!+\! z)]\!=\!\pm 0.4$.
E.g., for a central value $z\sim 1.95$ the uncertainty range is 
$0.18\,\lsim\, z\,\lsim\, 6.5$. Without understanding the origin 
of the spread, one cannot pin-down individual
redshifts from this correlation, or a cumulation of similar ones.
 
\begin{figure}[t]
\vskip -1cm
\begin{center}
\vbox{\epsfig{file=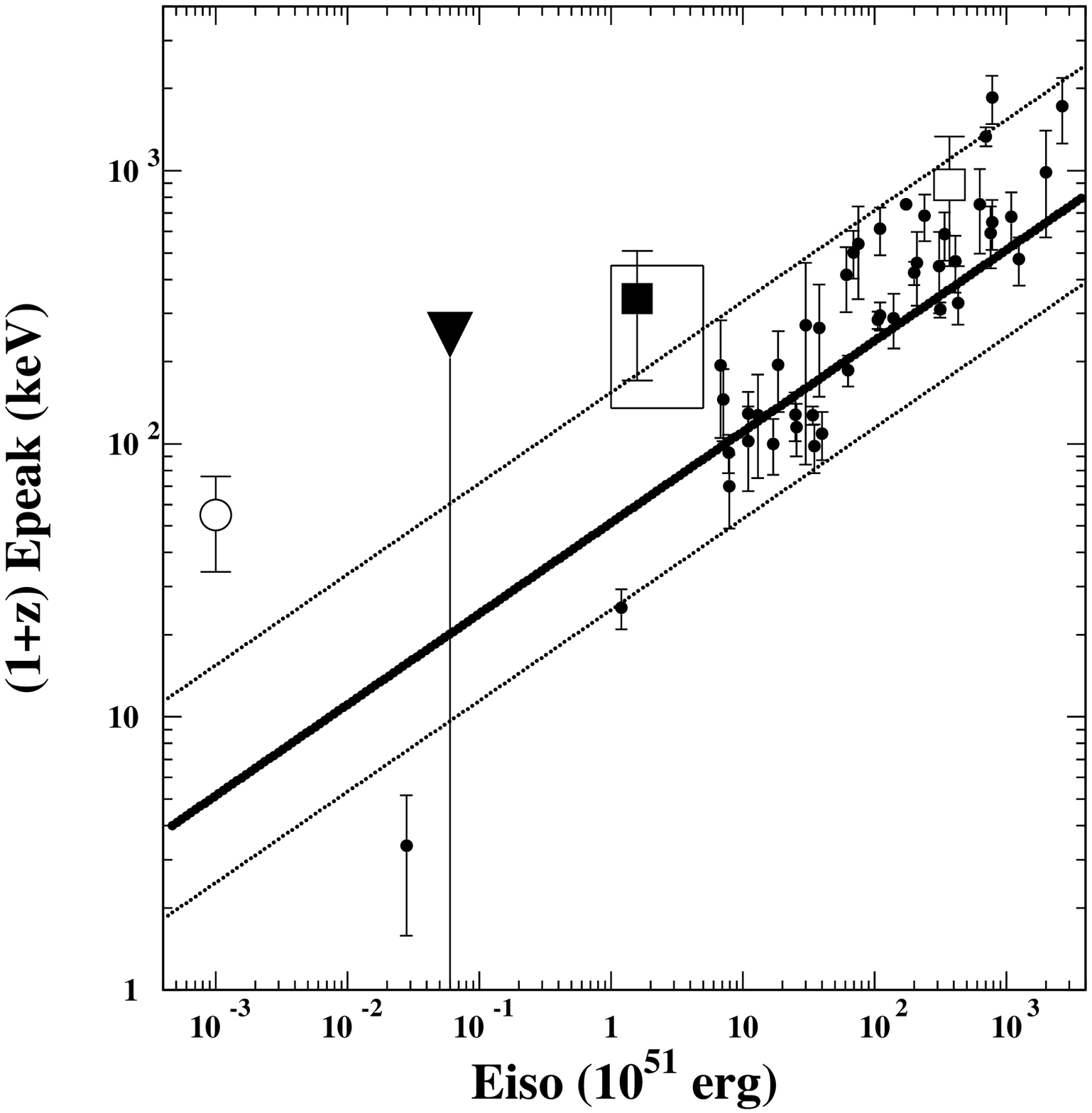, width=10cm}
\epsfig{file=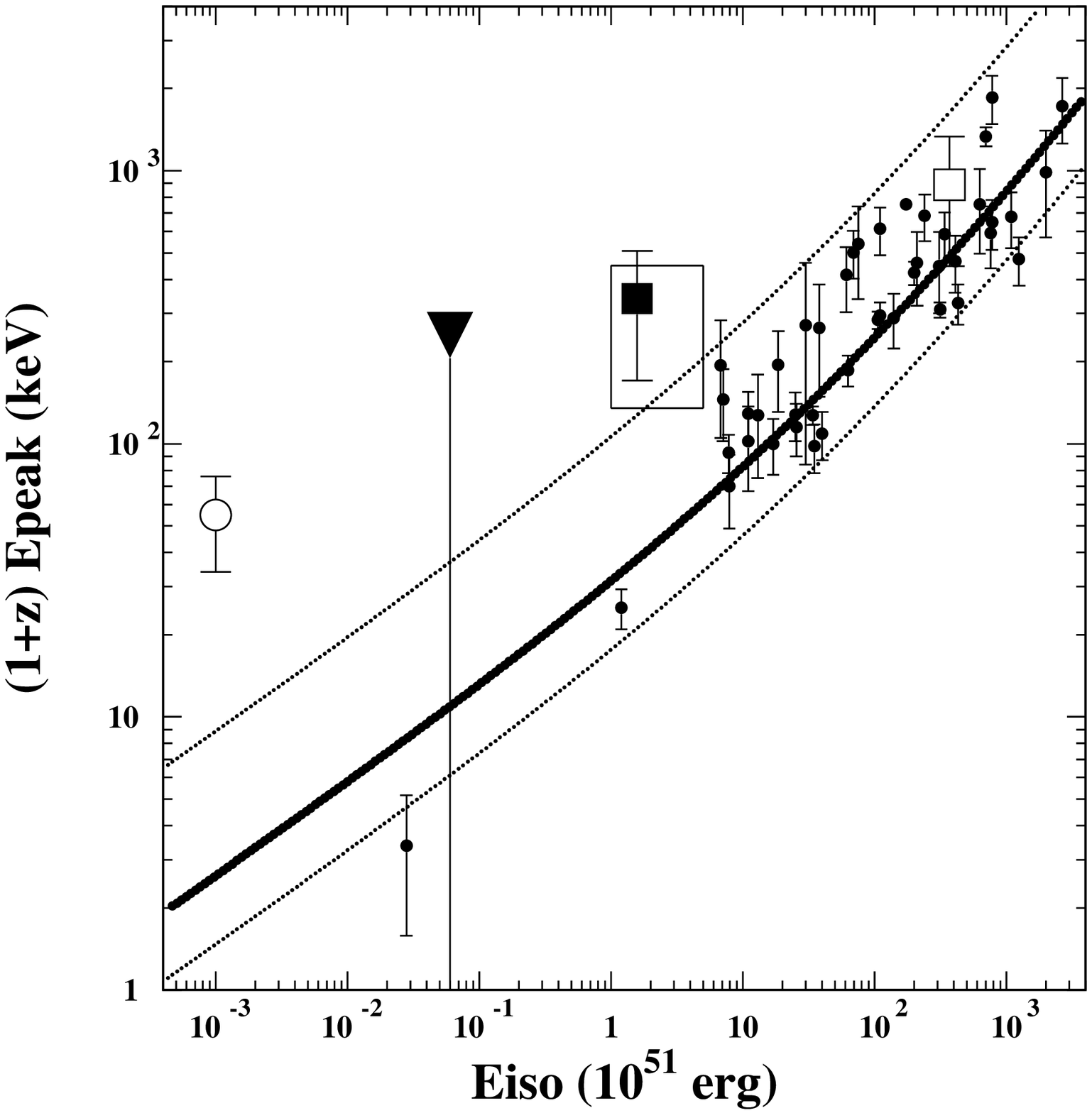, width=10cm}}
\caption{ $(1\!+\! z)\, E_p$ as function of $E_\gamma^{\rm iso}$ for 
a sample of 46 GRBs 
with secured redshift, 
compiled by Amati~(2006) and Ghirlanda et al.~(2004). 
 The rectangle is the CB-model's expectation
for a very faint SN at $z\!=0.125$.
{\bf (a): Top.} Our `pre-Amati' prediction.
{\bf (b): Bottom.} the improvement of  Eq.~(\ref{epi}).
The `variability' lines are the lightly-dotted ones.
GRB060614, for $z\!=1.95$ (0.125) is
the open (filled) square.
The open circle (GRB 0980425)
is convincingly an outlier. A CB-model's explanation is discussed in
Dado \& Dar (2005).}
  \label{fig1}
\end{center}
\end{figure}

A `pre-Amati' correlation was predicted [and tested]
in Dar \& De R\'ujula (2000b, [2004]).
According to Eqs.~(\ref{eobs},\ref{eiso}), 
$(1\!+\! z)\,E_p\!\propto\!\gamma\delta$
and $E_\gamma^{\rm iso}\propto\delta^3$. If most of the variability
is attributed to the  fast-varying $\theta$-dependence of $\delta$ in
Eq.~(\ref{delta}),
$(1\!+\! z) E_p\!\propto\![E_\gamma^{\rm iso}]^{1/3}$. 
This prediction 
is compared to current data in Fig.~\ref{fig1}a
(the `variability lines'
are not symmetric about the best-fit, because most
data have similar relative errors: the lower-$E_p$ ones have
smaller absolute errors and `attract' the best-fit line).
The agreement can be further improved 
by exploiting another prediction. A typical observer's
angle is $\theta\!\sim\!1/\gamma$. A relatively large $E_p$
implies a relatively large $\delta$, and a relatively small viewing angle,
$\theta<1/\gamma$. For $\theta^2 \ll 1/\gamma^2$, 
$\delta\simeq 2\gamma$, implying that $(1\!+\! z)E_p\propto 
[E_\gamma^{\rm iso}]^{2/3}$ {\it for the largest observed values} of 
$E_\gamma^{\rm iso}$. On the other hand, for $\theta^2 \gg 1/\gamma^2$,
the `pre-Amati' correlation is unchanged: it should be increasingly accurate 
{\it for smaller values} of $E_\gamma^{\rm iso}$.
We interpolate between these extremes by positing:
\begin{equation}
 (1\!+\! z)\,E_p=A\,[E_\gamma^{\rm iso}]^{1/3}+B\,[E_\gamma^{\rm iso}]^{2/3} \; .
\label{epi}
\end{equation}
A best fit to Eq.~(\ref{epi}) is shown in Fig.~\ref{fig1}b, an 
a-posteriori improvement over Fig.~\ref{fig1}a. The variability 
is due  to potentially varying intrinsic parameters. In Eq.~(\ref{eiso}),
for instance, there are four of them, besides $\delta$. The fit to
Eq.~(\ref{epi}) has $\chi^2/{\rm dof}\!=\!11.4$, similar to that
of Amati's arbitrary-power correlation (11.7). Yet, the correlations are not 
reliable estimators for the redshift of {\it individual} GRBs: in Figs.~\ref{fig1},
GRB060614 at $z\!=0.125$ is not a convincing outlier.  GRB060614
would not be an outlier, had the `variability line'
encompassed GRB980425 (a maverick outlier, but for an allegedly good
reason, see Dado \& Dar 2005).

\begin{figure}[t]
\vskip -1.cm
\begin{center}
\vbox{\epsfig{file=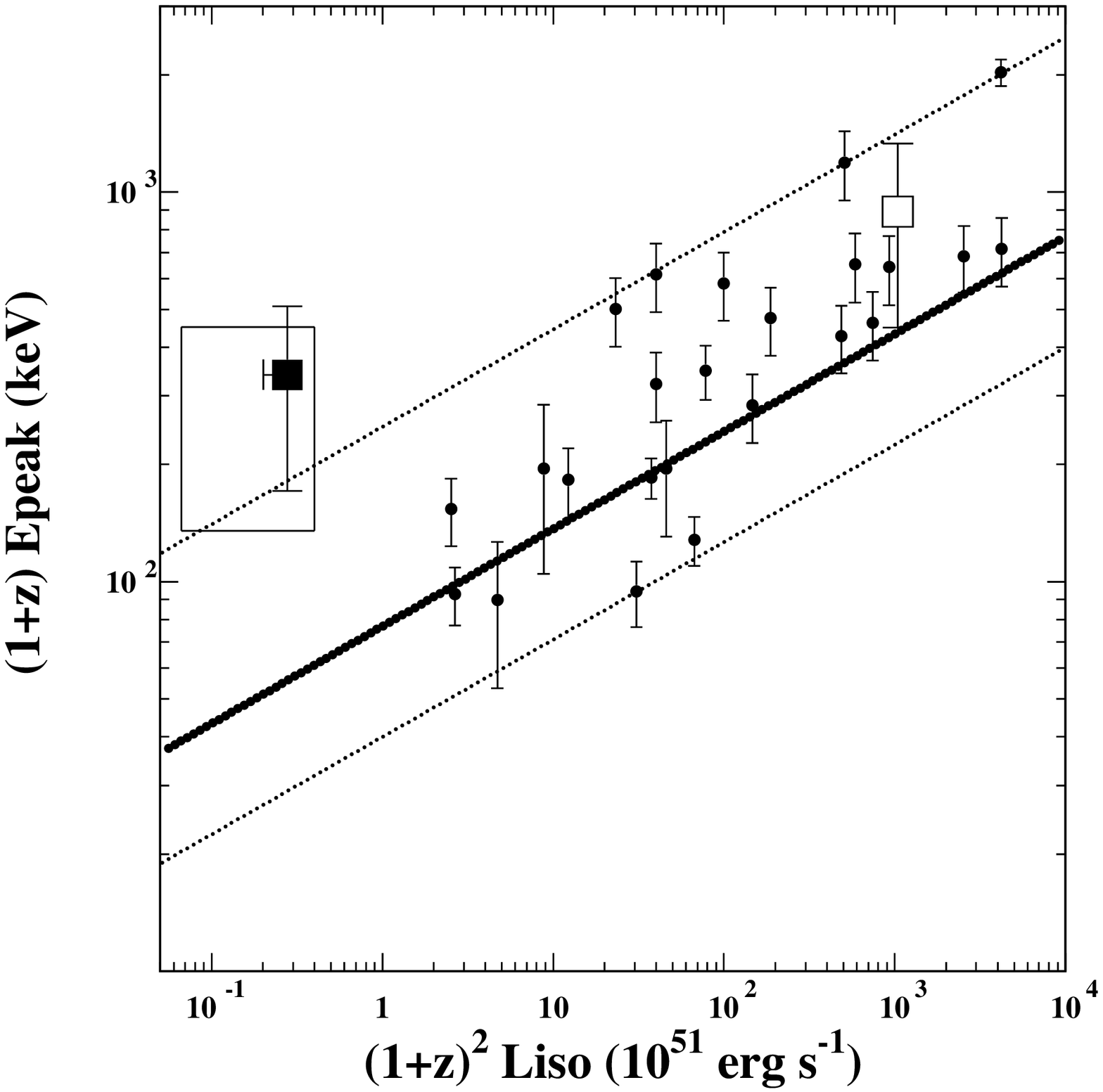, width=10cm}
\epsfig{file=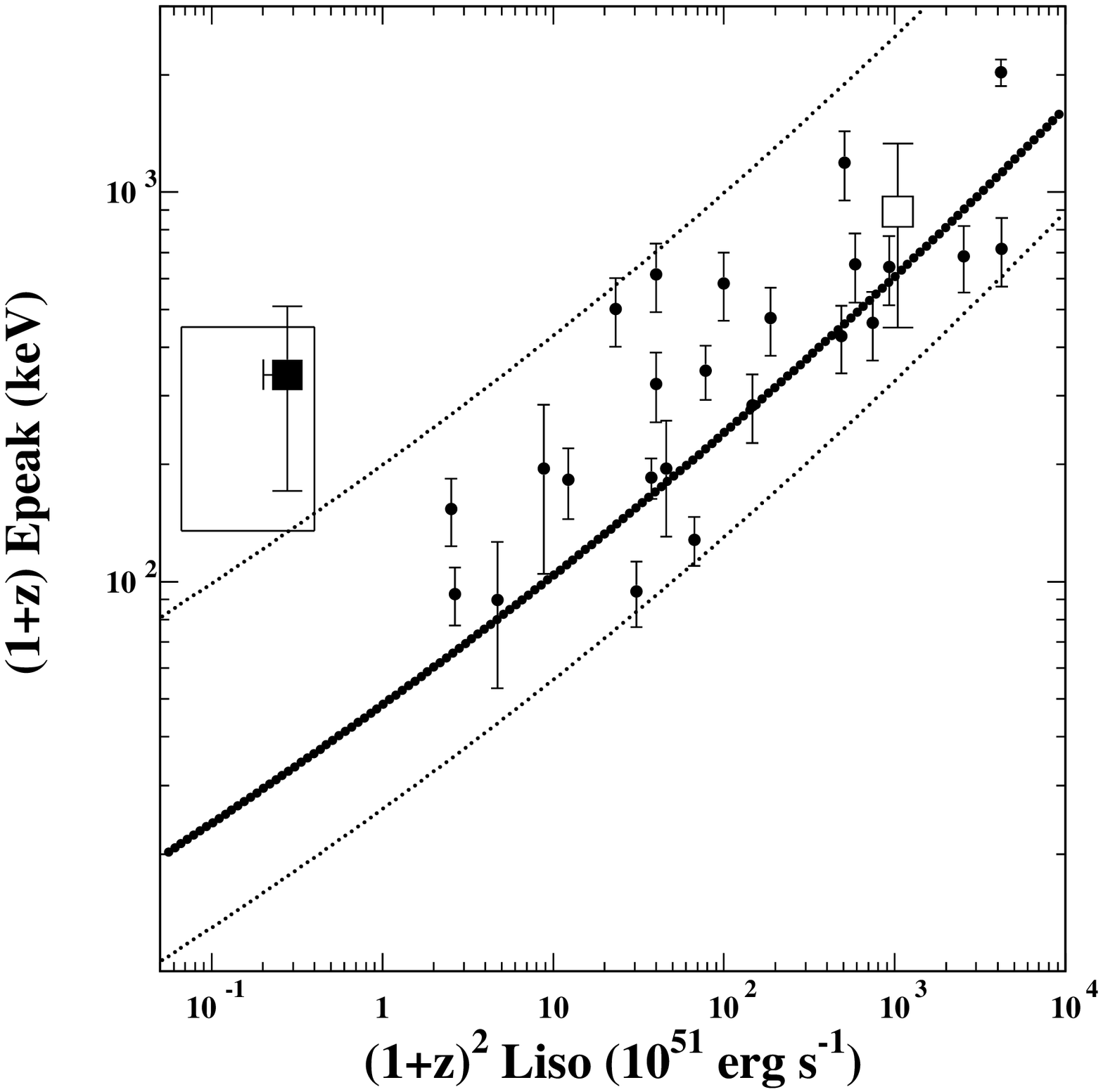, width=10cm}}
\caption{
 $(1\!+\!z)E_p$ as function of $(1\!+\!z)^2L_p^{\rm iso}$ for 
a sample of GRBs
with secured redshift, 
compiled by Yonetoku et al.~(2004) and Ghirlanda et al.~(2005). 
The rectangle is the CB-model's expectation
for a very faint parent SN at $z\!=0.125$.
{\bf (a): Top.} Our `pre-Yonetoku' prediction.
{\bf (b): Bottom.} The improvement of  Eq.~(\ref{lpep}).
Notation and comments are as in Fig.~\ref{fig1}.}
  \label{fig2}
\end{center}
\end{figure}

Another estimator is based on the 
correlation $(1\!+\! z)\, E_p\!\propto\! [L_p^{\rm iso}]^{0.51}$ 
(Yonetoku et al.~2004).
Paraphrase our discussion of the $[(1\!+\! z)E_p,E_\gamma^{\rm iso}]$ case,
using Eqs.~(\ref{delta}-\ref{lpeak}), 
to find that $(1\!+\! z)E_p\!\propto\![(1\!+\! z)^2 L_p^{\rm iso}]^{c}$,
with $c\!=\!1/4\,(1/2)$ for small (large) $L_p^{\rm iso}$. In Fig.~\ref{fig2}a we test
our `pre-Yonetoku' prediction ($c\!=\!1/4$, Dar \& De R\'ujula 2004). 
In Fig.~\ref{fig2}b, the prediction is improved, positing:
\begin{equation} 
(1+z)\,E_p \simeq C\, [(1+z)^2\,L_p^{\rm iso}]^{1/4} + 
D\, [(1+z)^2\,L_p^{\rm iso}]^{1/2}  \; .
\label{lpep}  
\end{equation}
The corresponding fit has $\chi^2/{\rm dof}\!=\! 6.8$; for Yonetoku's relation 
it is 8.0. Once more, the variability is too large to pin-down  the redshift of 
GRB060614. 

Some redshift estimators are pre-improved by employing Frail's\footnote{ 
The Frail relation (Frail et al.~2001), though used extensively
in the literature, has a trivial geometrical error. 
It should read
$E_\gamma=E_\gamma^{\rm iso}(1-\cos\theta_j)/2\!\simeq\!
E_\gamma^{\rm iso}\theta_j^2/4$.} 
`true' GRB
energy, $E_\gamma$, 
and the ensuing `true' luminosity, $L_p$,
in the correlations, e.g.~[$(1\!+\!z)E_p\!-L_p$] (Ghirlanda et al.~2005)
and  [$(1\!+\!z)E_p\!-E_\gamma$]  (Schaefer \& Xiao 2006).
This procedure may be unreliable: 

\noindent
1) Even if GRBs were produced by conical ejecta, the opening angle, $\theta_j$, 
of the jet during the GRB and AG phases may not be the same. 
Analogous jets from quasars and microquasars are not conical shells, but
plasmoids (CBs) whose rapid expansion stops shortly after ejection 
(Dar \& De R\'ujula 2004 and references therein).
Their radiation is beamed into a narrow cone, not a good reason to spouse 
conical jets. Moreover, the
CBs of quasars (Sambruna et al.~2006) and microquasars
(Namiki et al.~2003 and references therein) appear to be made of ordinary-matter
plasma (Dar \& De R\'ujula 2000a,2004) and not of $e^+e^-$ pairs.

\noindent
2)  The break in the AG, argued to occur 
when the observer begins to see the full front of the 
conical jet, must be achromatic, but it is not (e.g.~Panaitescu et al.~2006).

\noindent
3) The break time depends not only on $\theta_j$ and $E_\gamma^{\rm iso}$ 
but also on the chosen circumburst density distribution
(a constant, or the $\sim\!1/r^2$ profile of the wind of a Wolf-Rayet progenitor), 
on its normalization, and on the efficiency for converting  kinetic 
energy to radiation. These `hidden variables' may on occasion be chosen 
to converge on the desired result: a fixed `true' energy. If so, it is not surprising
that the ensuing correlations appear to be tighter.

\noindent
4) For all XRFs of known $z$, $E_\gamma^{\rm iso}$ 
is much smaller than the  Frail `standard candle' value $E_\gamma$,
implying that XRFs cannot be simply GRBs viewed far 
off axis, while all the observations support that they are, including  the predicted
(Dar \& De R\'ujula 2000a, 2004; Dado et al.~2002, 2003, 2004) 
and observed (Pian et al.~2006) SN1993bw-like 
progenitors\footnote{The observers claim that the data
``{\it suggest} that XRF 060218 is an intrinsically weak and soft event, 
rather than a classical GRB observed off-axis.'' In Dado et al.~(2007b)
we prove the contrary.}.

\noindent
5) The Frail relation and most of its consequences are derived for
observers placed on the firecone's axis, to within a beaming
angle $\sim 2/\gamma$. The ratio of the probability of being on-axis
to that of being `on-edge' (to within the same angle) is $\theta_j/\gamma$.
The on-axis/off-axis probability ratio is quadratic in $\theta_j/\gamma$. For
typical firecone parameters these probability ratios are tiny.

\noindent
6) All published attempts to {\it predict} the AG's break time
of a GRB, using the measured values of $z$,
$E_p$ and $E_\gamma^{\rm iso}$, have failed. For instance, 
Rhoads et al.~(2003) predicted $t_{\rm break}>10.8$ days for GRB 030226, 
while Greiner et al.~(2003), shortly after, observed $t_{\rm break}\sim 0.8$ 
day.

In view of the above, it is not surprising that SWIFT  X-ray and optical 
data, and ground-based observations of the corresponding optical AGs,
do not support the fireball model interpretation of the light-curve
`breaks' (e.g., Burrows \&
Racusin~2006;  Liang et al.~2007), nor the Frail relation (e.g., Kocevski
\& Butler~2007).

\section{Conclusions}

It has been stated that the deep limit
--at  $z\!=\!0.125$-- on a supernova associated with GRB 060614
constitutes the discovery of a new-object class
(Gal-Yam et al.~2006; Della Valle et al.~2006; Fynbo et al.~2006;
Gehrels et al.~2006).
In spite of the limited statistics, the conclusion may be correct,
and the new object may even be one of the ones previously
discussed in the literature: a `failed supernova'
(a direct collapse 
into a black hole with no visible SN, Woosley 1993), a supranova
(a delayed collapse of a neutron star, Vietri \& Stella 1998), or 
a phase transition between the possible states of a compact 
hadronic star (Shaviv \& Dar~1995).  All the properties of  
GRB 060614, but the lack of a SN progenitor, are compatible with those
of typical long-duration GRBs.
If this GRB belongs to an entirely new class, this would need to be
explained.

We have discussed less drastic conclusions, and studied
three reasons why a SN progenitor of GRB 060614 may have 
avoided detection: strong extinction of the SN light in a molecular 
cloud, a fake sky coincidence with a  galaxy at $z\!=\!0.125$,
and a dimmed associated  supernova. The merits and demerits of these 
mutually-excluding alternatives are:
\begin{itemize}
\item
 The molecular-cloud hypothesis is the most obvious. 
In fireball models it would be excluded by the available data
on the afterglow's optical extinction. But in the CB model,
in which the SN and the source of the afterglow (the moving 
cannonballs) do not lie in the same fixed direction, it is not
 excluded. We disfavoured it on grounds that
it cannot naturally accommodate the values of all the prompt
GRB observables.
\item
The possibility that the GRB is much
more distant than $z\!=\!0.125$ (Schaefer and Xiao, 2006; 
Cobb et al.~2006) is  consistent with the data, but not decisively 
provable. The correlations between GRB observables
used to substantiate this hypothesis are, as we have
discussed in detail, not sufficiently trustable for a single case.
This is so even in the CB-model, wherein the correlations 
satisfied by the data are predictions based on trivial physics 
and mere geometry (Dado, Dar \& De R\'ujula, 2007a and references
therein). To agree with all observations, this scenario
requires the GRB to be host-less, or to
have taken place in a dwarf galaxy fainter than 
the HST limit (Gal-Yam et 
al.~2006); that the line of sight to the GRB not include significant 
Lyman break absorbers, in order to be consistent with the SWIFT/UVOT 
results (Gehrels et al.~2006); and that the column density of dust
be low along the line of sight
in the observed dwarf galaxy at $z\!=\!0.125$, as inferred by 
Gal-Yam et al.~(2006) from their spectral analysis.

\item
The proposal that GRB 060614 did occur at $z\!=\!0.125$ and
was associated with a very faint
supernova is the most economical. It adopts the most probable
redshift. It maintains the established 
association between long-duration GRBs and SNe. It appeals
to a type of SN which is known to exist (Turatto, 1998;
Pastorello et al.~2004, 2007).
In the CB model this alternative is supported by the fact that it
results in correct predictions for the observed peak energy,
isotropic energy and peak luminosity of the GRB. 

\end{itemize}

Two out of the five secured GRB/SN associations,  
(GRB021211/SN2002lt: Della Valle et al.~2003; 
GRB060218/SN2006aj: Pian et al.~2006;
Mazzali et al.~2006),
involved SNe significantly different from SN1998bw. Recently, Li (2007) 
argued that the peak energy, $E_p$, 
of their prompt $\gamma$-ray emission  
is correlated to the peak bolometric luminosity 
$L_p(SN)$ of their SN:
$E_p\propto [L_p(SN)]^{4.97}$. If such a correlation
were true, it would exclude both a faint and a `failed supernova' origin 
of GRB 060614. But the alleged correlation is highly uncertain:
it is based on four events --three of them clustered within errors--
and on measurements with large
systematic uncertainties. Moreover, it
has  not been tested for the many more
GRB/SN pairs with a photometrically discovered SN. 

After decades of intense theoretical efforts
and numerical simulations, there is no first-principle
understanding of SN explosions, not to speak of the allegedly consequent
generation of relativistic jets, particularly if required to consist
mainly of $e^+e^-$ pairs. Yet, most fireball-inspired discussions of 
the association advocate SNe Ib/c. 

The hypotheses of the CB model do not stem from simulations, but from
the analogy between observations of accreting quasars or microquasars,
with expectations --based on angular-momentum conservation-- about
the inner realms of a rotating star whose core has collapsed.
It is not obvious why the external composition of the star, which
determines the SN type, would play a crucial role. Thus the model
does not imply a strong a-priori preference for a SN type, provided it
is (for long GRBs) a core-collapse SN. Evidence that Type II
SNe emit CBs is provided by the ``mystery spots" of SN1987A
(Ninenson \& Papaliolios, 2001). 

There are large uncertainties in the conversion of the measured rates of
GRBs into the true frequency of these events. They arise not only from
detection biases, backgrounds and thresholds, but from the model-dependent
beaming-angle distribution. Moreover, in the CB model, a GRB's luminosity
is proportional to the early luminosity of its generating SN, which is
type-dependent. And so is the SN luminosity function. All in all, for the
typical very narrow beaming angles of the CB model, $\theta\!\sim\!1$
mrad, the uncertainties are large enough to amply accommodate the
conclusion that most core-collapse SNe generate GRBs, or that only
$\sim\!15$\% of them do it, perhaps the ones of type Ib/c.

Gal-Yam et al.~(2006b) and 
Soderberg  et al.~(2006) are often quoted for their 1\% upper
limit on the fraction of GRB-generating SNe. This ``observation" relies
entirely on the fireball model and the Frail relation,
which are not supported by the bulk of the SWIFT data
(see, e.g., Kumar et al.~2007;
Burrows \& Racusin~2007; Kocevski \& Butler~2007: Urata et al.~2007;
Zhang, Liang \& Zhang~2007; Yonetoku et al.~2007; Liang et al.~2007).
Moreover, Soderberg et al.~(2006) used data that are not verified. Some of 
their ``bright" SNeIc  were even thought to be possible SNeIa by the 
original observers. 

On the other hand, except for GRB060218/SN2006aj 
and GRB021211/SN2002lt, the three other  well observed GRB-associated SNe
were similar but much brighter than ordinary
SNeIb/c (e.g., Mazzali et al.~2006), perhaps because they were seen so
close to their GRB's jet axis. Podialkowski et al.~(2004)
estimated that the rate of such bright SNeIc  is consistent
with the rate of long GRBs, but their estimated GRB rate was also based on
a beaming angle extracted from the Frail relation.

In the CB model the GRB-generating SNe need not be `hypernovae',
but ordinary SNe viewed close to their axis, a direction in which their
non-relativistic ejecta may be faster than average, 
as observed. This expectation of non-exceptionality 
appears to be corroborated by the observation
that SN2006aj was quite different from SN1998bw, SN2003dh and SN2003lw 
(Pian et al.~2006; Mazzali et al.~2006).
For a detailed discussion of the SN/GRB association,
see Dar \& De R\'ujula (2004).

We have argued that a SN may well have been associated with GRB060614,
either a 1998bw-like SN at $z\!\sim\!2$, or a two-three orders of 
magnitude less luminous SN, at $z\!\sim\!0.125$.
What type of SN was it? Since it was not observed, the issue
cannot be addressed very decisively. The 
claim that GRB060614 belongs to a new 
class of GRBs with no associated supernova  will also remain 
unratified until many other such cases are observed.

\noindent
{\bf Acknowledgments.}
{ We are indebted to Avishay Gal-Yam
for very useful discussions and to an anonymous referee for useful 
comments and suggestions.
This research was supported in part by the
Helen Asher Space Research Fund at the Technion Institute. }

\pagebreak

\pagebreak

\end{document}